\documentclass[preprint,3p,times,twocolumn]{elsarticle}
\usepackage{graphicx}
\usepackage{color}
\usepackage{hyperref}
\usepackage{amssymb}
\usepackage[figuresright]{rotating}

\journal{Computer Physics Communications}

\begin{document}
\begin{frontmatter}

\title{Using zeros of the canonical partition function map to detect signatures of a Berezinskii-Kosterlitz-Thouless transition}
\author[ufmg,ufop]{J.C.S. Rocha\corref{correspondingauthor}}
\cortext[correspondingauthor]{Corresponding author}
\ead{jcsrocha@iceb.ufop.br}
\author[ufmg]{L.A.S. M\'ol}
\ead{lucasmol@fisica.ufmg.br}
\author[ufmg]{B.V. Costa}
\ead{bvc@fisica.ufmg.br}
 \address[ufmg]{ Laborat\'orio de Simula\c c\~ao, Departamento de F\'isica, ICEx  Universidade Federal de Minas Gerais, 31720-901\\ Belo Horizonte, Minas Gerais, Brazil}
 \address[ufop]{Departamento de F\'isica, ICEB, Universidade Federal de Ouro Preto, 35400-000 Ouro~Preto, Minas Gerais, Brazil}

\begin{abstract}

         Using the two dimensional $XY-(S(O(3))$ model as a test case, we show that analysis of the Fisher zeros of the canonical partition function can provide signatures of a transition in the Berezinskii-Kosterlitz-Thouless ($BKT$) universality class. Studying the internal border of zeros in the complex temperature plane, we found a scenario in complete agreement with theoretical expectations which allow one to uniquely classify a phase transition as in the $BKT$ class of universality. We obtain $T_{BKT}$ in excellent accordance with previous results. A careful analysis of the behavior of the zeros for both regions $\mathfrak{Re}(T) \leq T_{BKT}$ and $\mathfrak{Re}(T) >  T_{BKT}$ in the thermodynamic limit show that $\mathfrak{Im}(T)$ goes to zero in the former  case and  is finite in the last one.
\end{abstract}
\begin{keyword}
{Phase transitions: general studies}\sep{Classical spin models}\sep{Monte Carlo methods}
\PACS{05.70.Fh}\sep{75.10.Hk}\sep{05.10.Ln}
\end{keyword}
\end{frontmatter}
\section{Introduction}
    The Berezinskii-Kosterlitz-Thouless ($BKT$) transition already has more than 40 years of history~\cite{40Years} and is still intriguing the scientific community. The nature of this transition is completely different from the common discontinuous (first order) or continuous (second order) phase transitions. Long range order does not exist and the two point correlation function has an algebraic decay at low temperature ($T \leq T_{BKT}$) and an exponential decay for $T > T_{BKT}$~\cite{Mermin_Wagner}. Here $T_{BKT}$ is known as the $BKT$ temperature, and a model displaying a $BKT$ transition has an entire line of critical points for $T \leq T_{BKT}$. In addition to that, the corresponding free energy, which is a $C^\infty$ function, is not analytical in this region. Besides these striking features, the correlation function has a characteristic universal exponent decay $\eta(T_{BKT}) = 1/4$ at the transition temperature. Its phenomenology relies on the belief that it is driven by a vortex-antivortex unbinding mechanism~\cite{B,KT}. Another proposition that also describes the transition is based on a polymerization of domain walls~\cite{Patrascioiu}. Many systems, e.g., superfluid films, Coulomb gases and crystal surface roughening, undergo transitions that can be classified as belonging to this universality class~\cite{Minhagen}. More recently, Berker et al~\cite{Berker} found that a  $BKT$ transition can occur in a scale free network.

    Although the $BKT$ transition is well known, the characterization of an unknown phase transition as being in the $BKT$ universality class is not an easy task since there is no standard method to do so. The lack of a criterion capable of determining the nature of a transition beyond any reasonable doubt  is a problem discussed by Bramwell and Holdsworth~\cite{Bramwell_Holdsworth}. They pointed out that to be able to see the transition the system under investigation  should be very large. They estimate that for ``a system with atomic spacing of $ 3{\AA}$ the area should correspond to the size of a postage stamp''. From the analytical point of view, the renormalization group approach is able to describe the main features associated with the transition. However, the approximations involved in the study of a given model may hinder the discovery of its real nature. 
    
   In this paper we present an algorithm for the systematic analysis of the Fisher zeros of the canonical partition function~\cite{YL,LY,Fisher_1} for the 2D XY model (with spin symmetry $O(3)$) looking for possible signatures of the $BKT$ transition. It was Fisher~\cite{Fisher_1}, in $1964$, who proposed considering the partition function zeros in the \emph{complex temperature plane} to study phase transitions. As known, the thermodynamic behavior of a given physical system is encoded by its partition function, $Z$, and all thermodynamic quantities can be obtained as derivatives of the free energy, $F=-k_BT \ln{Z}$. The basic assumption of the Fisher zeros approach to study phase transitions is that a system undergoes a phase transition at a given (real) temperature, $T_c$, if $Z(T_c)=0$, reflecting the non-analyticity of $F$ at $T_c$~\cite{Wei_1}. Since in a $BKT$ transition there is an entire line of critical points for $T\leq T_{BKT}$, one should expect that a map of the Fisher zeros in the complex temperature plane exhibits an entire line of zeros in the real temperature axis for $T \leq T_{BKT}$, signaling the $BKT$ behavior of the transition, while for $T>T_{BKT}$ the zeros should not touch the real positive axis, emphasizing the analytical behavior of thermodynamic functions at high temperatures. Of course, these considerations apply only to the thermodynamic limit. We warn the reader that the Fisher zeros studied here should not be confused with the Yang-Lee zeros~\cite{YL,LY} defined on the complex fugacity plane instead of the temperature plane~\cite{Kenna_1,Irving,Kenna_2}. In what follows we will analyze the Fisher zeros map for the XY-model.
\section{Model and Methods}
    Here we study a ``fruit fly'' model of the $BKT$ transition~\cite{Minhagen}, the classical two-dimensional XY-model on a square lattice, defined by
\begin{equation}
    \mathcal{H} = -J \sum_{\langle i,j \rangle} (S_i^xS_j^x + S_i^yS_j^y).
\end{equation}
\noindent
    The sum runs over the nearest neighbors, $J$ stands for the exchange coupling constant and $S_i^{\alpha}$ stands for the component ${\alpha}=(x,y,z)$ of the $i^{th}$ spin. The same Hamiltonian also defines the Planar-Rotator ($O(2)$) model~\cite{PRM1,PRM2}, whose spins have only two components (one degree of freedom), and can also be viewed as an example belonging to the $BKT$ universality class. In spite of the lack of long-range order for the model~\cite{Mermin_Wagner}, there is a \emph{non-zero magnetization} for any finite volume~\cite{PLA_Costa}, resulting in a thermodynamic behavior very similar to that observed at continuous phase transitions. In fact, the behavior can be easily confused with a second order phase transition or something else~\cite{MC1,MC2,MC3}. Thus, distinguishing between continuous and $BKT$ behavior in finite systems is difficult and may require system sizes beyond those feasible. Usually, the $BKT$ temperature is estimated by using the Binder cumulant, the divergence of the magnetic susceptibility, the correlation functions or the more reliable helicity modulus~\cite{Minhagen,PLA_Costa}, quantities that do not assure the model is in the $BKT$ universality class.

    In the Fisher zeros approach one starts with a discrete canonical partition function that can be written as $Z=\sum_E g(E)\exp \left( {-\beta E}\right)$, where $g(E)$ stands for the density of states (DOS). For a continuous system one may perform a discretization of the DOS~\cite{Rocha_1}. This can be done by dividing the energy range into $M$ bins of size $\varepsilon$. We can organize the energies inside the interval $[E_0,E_{M-1}$] by counting the energy of the $n^{th}$ bin as $E_n = E_0 + n \varepsilon$. By defining a variable $x \equiv \mathrm{e}^{-\beta \varepsilon}$, we get a discretized version,
\begin{equation}
    Z_D =  \mathrm{e}^{-\beta E_0}\sum_{n=0}^{M-1} g_n x^n
    = \mathrm{e}^{-\beta E_0}\prod_{j=1}^{M-1} (x-x_j),
\end{equation}
\noindent
    where $g_n=g(E_n)$, and $x_j$ is the $j^{th}$ zero of the polynomial, usually called Fisher zero. Once $g_n$ is obtained, any reliable zero finder can be used \footnote{In this work we decided to use: E.W. Weisstein, ``Polynomial roots'', in MathWorld -- A Wolfram Web Resource, \url{http://mathworld.wolfram.com/PolynomialRoots.html}.} to find the $x_j$'s. As long as $g_n\in\mathbb{R}^+$, $Z$ is an analytical function and has no real positive zeros for finite systems. The zeros show up as complex conjugate pairs.
\noindent
\begin{figure}
      \includegraphics[width=0.4\textwidth,keepaspectratio=true,clip=true]{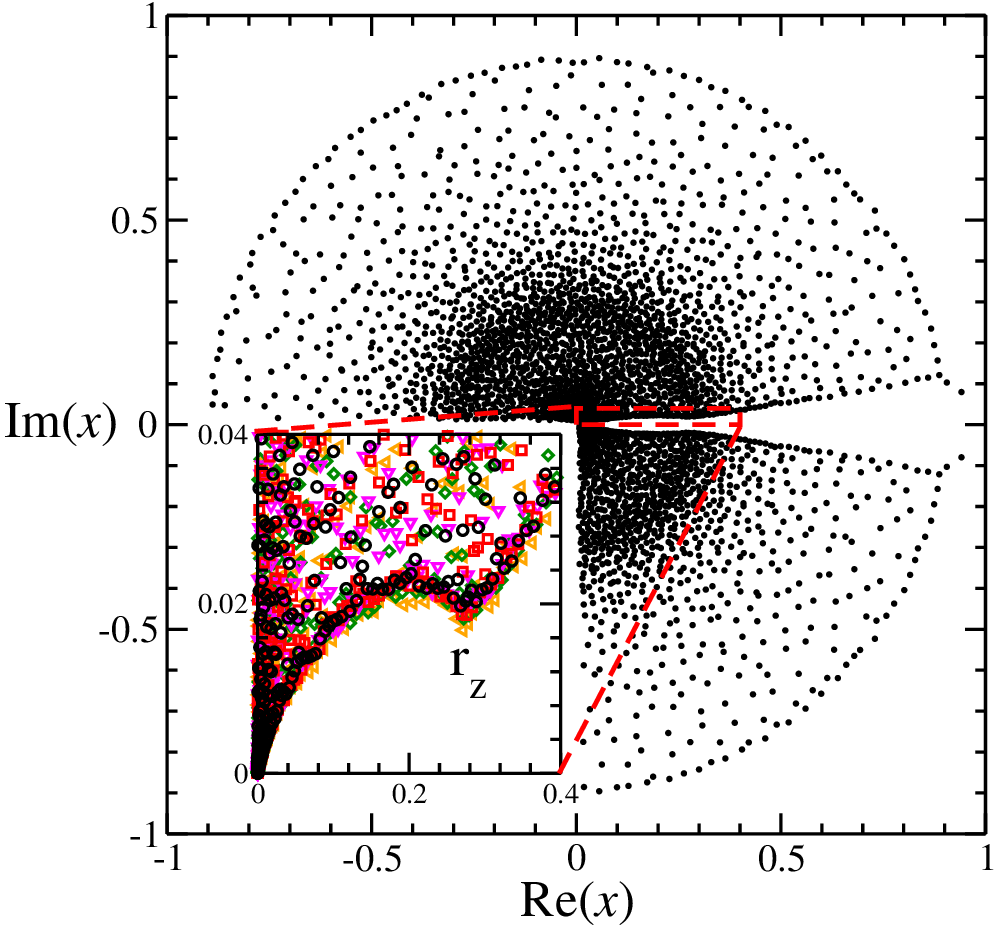}
    \caption{Fisher zeros map on the $x=\mathrm{e}^{-\beta \varepsilon}$ plane for the 2D classical XY-model in a $50\times50$ lattice. The inset shows a zoom on the inner region for five distinct simulations, represented by different symbols. The cusp, $r_z$, is indicated.
    \label{zeros_map_typical}}
\end{figure}
\noindent
    The analysis of Fisher zeros in finite systems is done by considering the special set of zeros $\{ x^\star(L) = a(L) + \mathrm{i} b(L) \} \in x_j(L)$, where $L$ is the linear size of the system, called first or leading zeros. They have the following properties: $b(L) \to 0$ as $L\to \infty$ while $\lim_{L \rightarrow \infty} a(L) = a(\infty)$, a constant value. The leading zeros are very stable against statistical fluctuations, in contrast to non-leading zeros, and are, in general, featured in the map (see, for example, Ref.~\cite{Rocha_2}).
    They are related to  the transition temperature of the system in the thermodynamic limit as $1/k_BT_c=-\ln (a(\infty))/\varepsilon$~\cite{Rocha_2} and their impact angle on the real positive axis is directly related to the order of the phase transition~\cite{Janke}. Recent results also suggest that the pattern of zeros as a whole can be used to characterize the order of the transition~\cite{taylor_1}. In a paper of $1983$ Ytzykson et al.~\cite{Itzykson}, analysing the zeros distribution for small lattices, found that some properties of the Ising and gauge models seem to exhibit a universal behavior close to complex singularities. However, it is not clear how to extend their arguments to the $BKT$ transition.
    
    In order to obtain the DOS of the XY-model we used the Replica Exchange Wang-Landau (REWL) method~\cite{Vogel_1,Vogel_2,Vogel_3,Vogel_4}, a parallel version of Wang-Landau (WL) sampling~\cite{wl1,wl2,wl3}, capable of sampling the entire configuration space efficiently in a single simulation. In this scheme the energy range is split into smaller overlapping windows. We considered $e_{0} = -1.9J$,  $e_{M-1} = 0$, and an overlap of $75\%$, where $e = E/L^2$ stands for the energy per spin. Several different random walkers are allowed to run in each of these windows following the original WL scheme\footnote{The considered flatness criteria is $p=0.7$, the final $\ln{f}$ value is $10^{-9}$, and the acceptance ratio is $60\%$.}. We use regular square lattices with sizes ranging from $L=10$ up to $200$. In this work we chose $J=1$, $S=1$, $k_B = 1$, and the lattice parameter $a=1$.

\section{Results and discussion}
    In Fig.~\ref{zeros_map_typical}  we show a typical zeros map of the imaginary and real parts of all $x_j$'s, for a  $50 \times 50$ lattice. In a continuous phase transition a single leading zero is expected. However, we cannot identify a single point featured in comparison to others. The inset shows in different symbols the zeros for five different simulations for $L=50$, in order to analyze statistical fluctuations. A leading zero cannot be identified in this picture. Instead, a cusp at $r_z$ is evident and the border line is quite stable against fluctuations.

\begin{figure}
    \includegraphics[width=0.45\textwidth,keepaspectratio=true,clip=true]{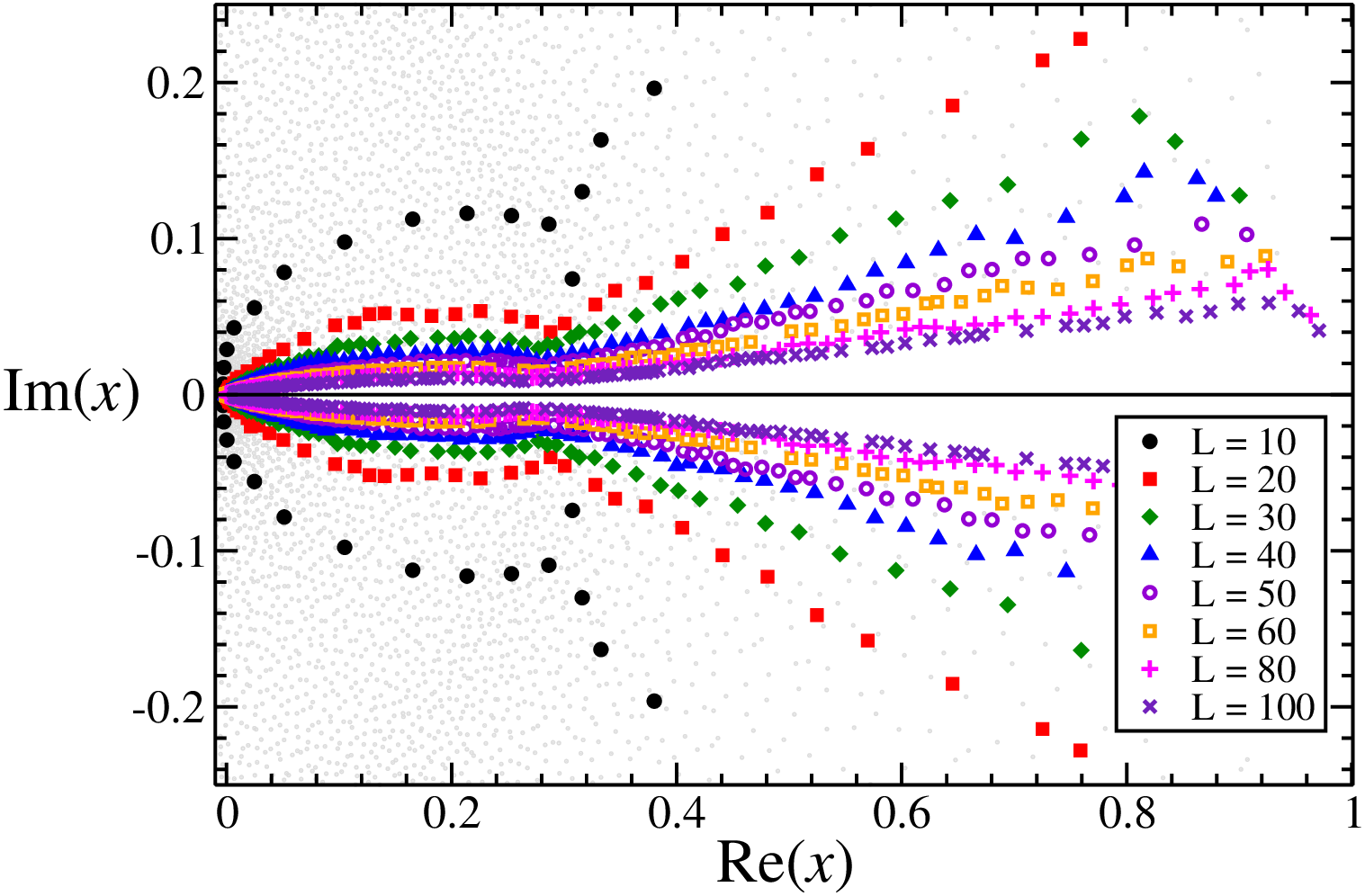}
    \caption{Zoom on the real positive semi-axis of the zeros maps in the $x$ plane for $L=10\to100$. The zeros on the internal border are highlighted.
    \label{zeros_map_zoom}}
\end{figure}

    The size dependence of the internal border of the map pattern as a function of the lattice size is shown in Fig.~\ref{zeros_map_zoom}. One should expect the internal border to coalesce into the real axis for $\mathfrak{Re}(x)\leq r_z$ and $L \to \infty$, in accordance with the existence of an entire line of critical points at low temperatures. The line for $\mathfrak{Re}(x) > r_z$ on the other hand should not touch the real axis. The cusp $r_z$ should, then, give $T_{BKT}$.
\noindent
\noindent

\begin{figure}
    \includegraphics[width=0.45\textwidth,keepaspectratio=true,clip=true]{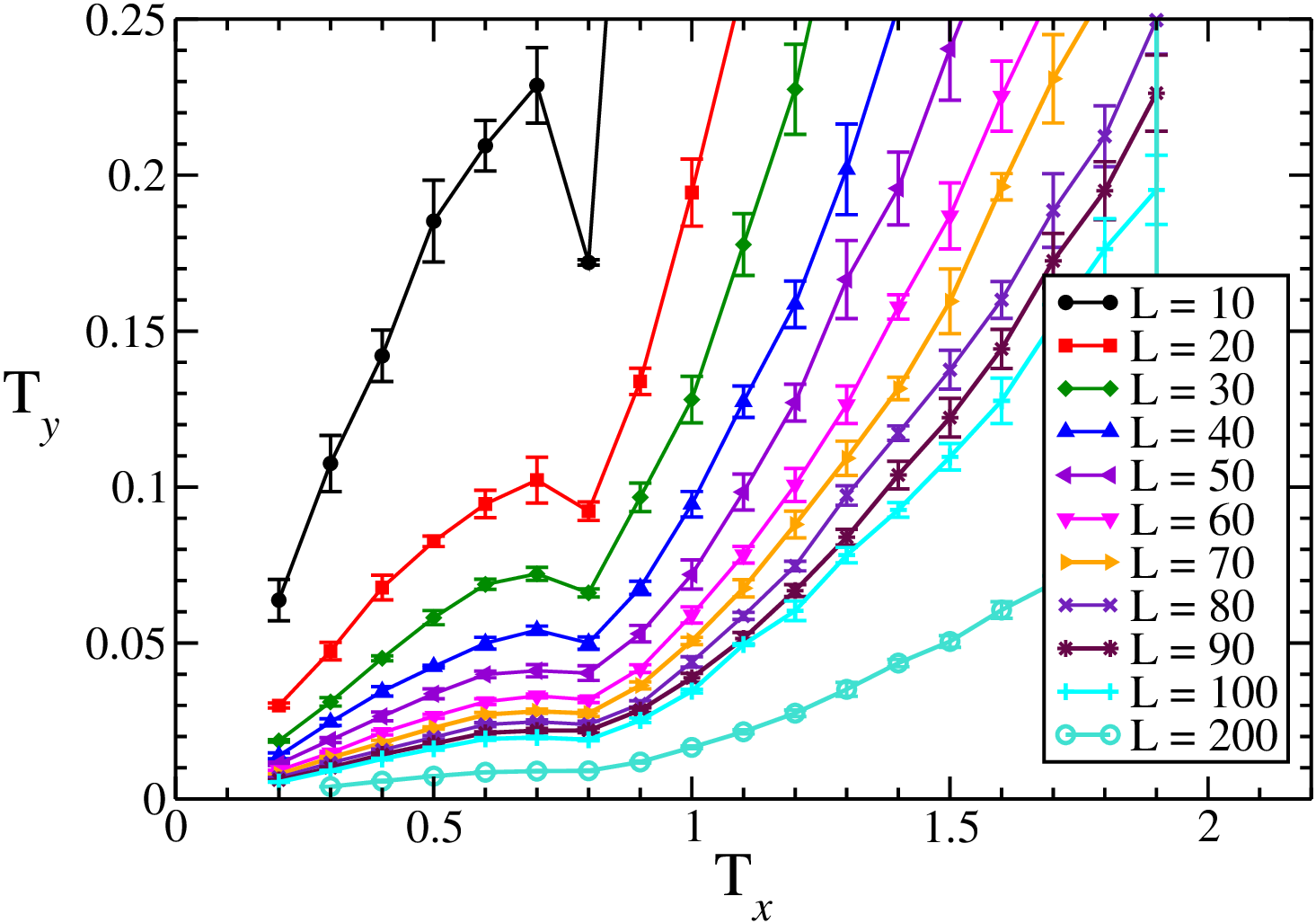}
    \caption{Internal border of zeros obtained by using the binning process for $L=10 \to 200$. Here $\Delta T_x =0.1$ and the error bars represent statistical fluctuations.
    \label{border_line}}
\end{figure}

\begin{figure}
    \includegraphics[width=0.45\textwidth,keepaspectratio=true,clip=true]{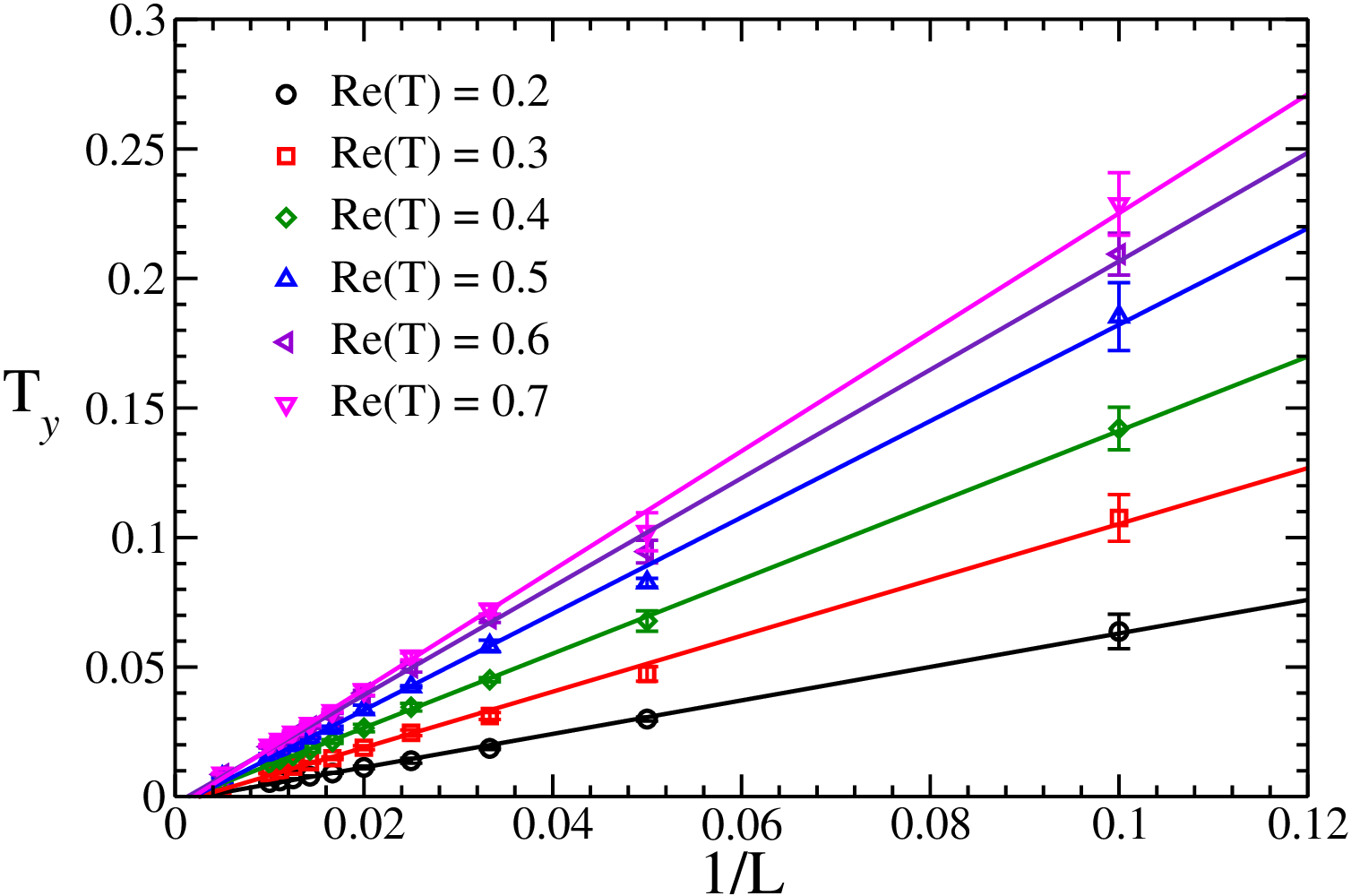}
        \caption{Finite size scaling analysis of the imaginary part of the internal border according to the \emph{ansatz} $T_y \sim L^{-1}$. The lines represent a linear regression of the data, showing good agreement with the \emph{ansatz} and the convergence to zero or small negative values for $L\to \infty$.  We use $\Delta T_x =0.1$.
    \label{fss_border}}  
\end{figure}

    In order to systematically investigate the finite size effects and confirm the above expectations, we opted to work in the complex temperature ($T$) plane instead of the complex $x=\mathrm{e}^{-\varepsilon/k_BT}$ plane. This choice is justified by the fact that finite size scaling in terms of the temperature is known~\cite{Kenna_2} and that by doing so our results are almost independent on the chosen bin size, $\varepsilon$. This last point is especially important when dealing with bigger lattices.
    In this work we used $\varepsilon = 1$ for $L\le100$ and $\varepsilon =3 $ for $L=200$; we noted that small modifications in the bin size did not change our results. To determine the limits of the internal border we divided the real temperature axis into small bins of size $\Delta T_x$ centered at $T_x=\mathfrak{Re}(T)$. The border line was identified by looking inside a given bin for the smallest value of $T_y=\mathfrak{Im}(T)$ that appears. At least five different simulations were used for each lattice size. The resulting curves are presented in Fig.~\ref{border_line}.
    Fig.~\ref{fss_border} shows $T_y$ as a function of $L^{-1}$ for some typical $T_x \le T_{BKT}$ values. The solid lines are linear regressions showing that $T_y \to 0$ for $L\to \infty$, i.e., the internal border coalesces with the real positive axis in the thermodynamic limit. Other scaling functions and exponents were also tested (not shown here), but they do not describe our data as well as this \emph{ansatz}.

\noindent

\begin{figure}[h!]
    \includegraphics[width=0.45\textwidth,keepaspectratio=true,clip=true]{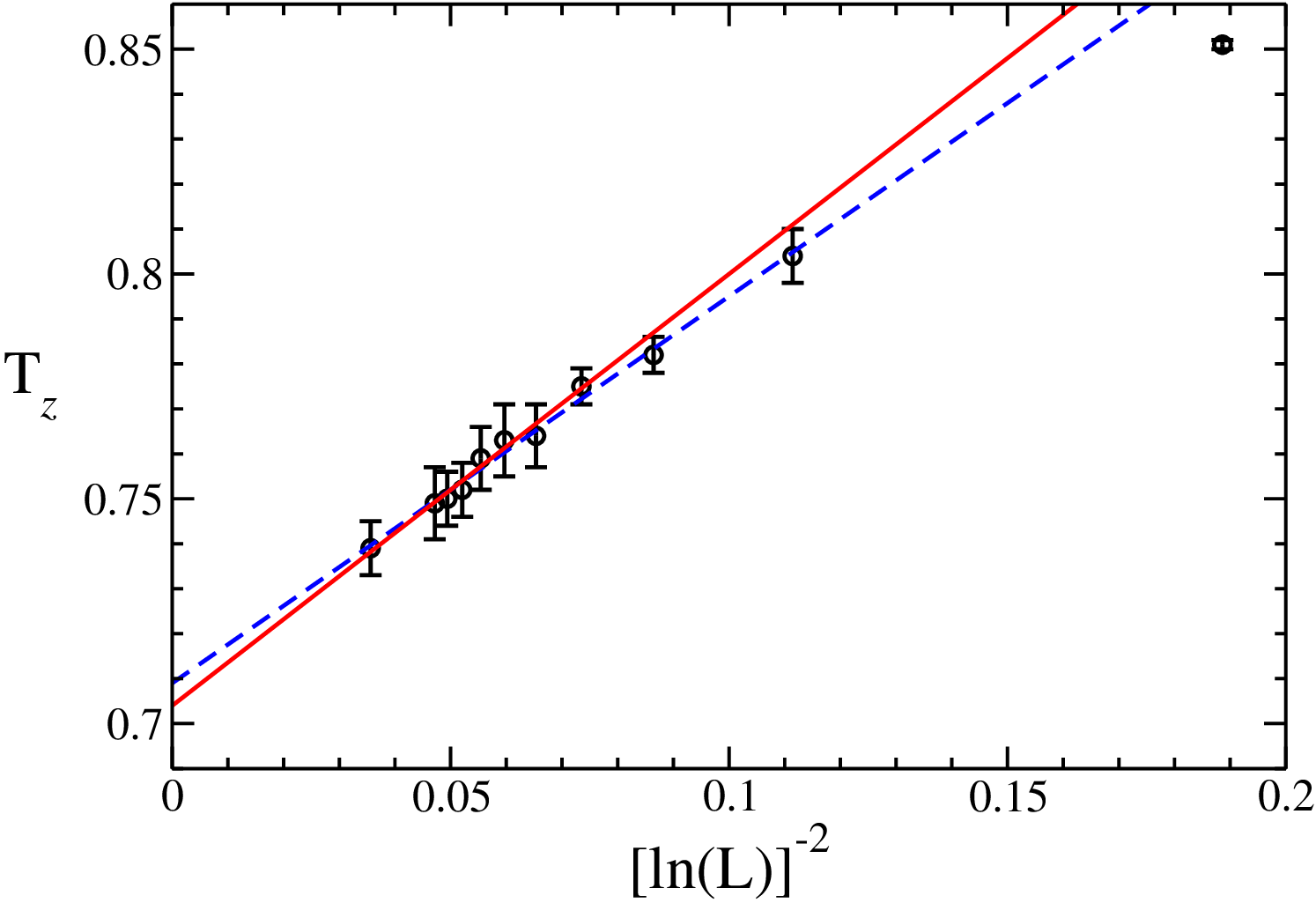}
    \caption{Finite size scaling of the cusp position, $T_z(L)$, according to the prediction for the $BKT$ ``pseudocritical" temperature $T_{BKT}(L) \sim [\ln(L)]^{-2}$. Discarding the point corresponding to $L=10$ the $BKT$ temperature is estimated by a linear regression as 0.709(2)  (dashed blue line) and discarding $L<40$, $T_{BKT}=0.704(3)$ (solid red line).
    \label{fss_T}}
\end{figure}

    To estimate the critical temperature we used the location of the cusp position, $T_z(L)$. Since the scaling function for the ``pseudocritical" temperature is given by~\cite{Kenna_2} $T_{BKT}(L) \sim [\ln(L)]^{-2}$ we plotted $T_z(L)$ as a function of  $[\ln(L)]^{-2}$ in the Fig.~\ref{fss_T}. A linear regression, discarding the point corresponding to $L=10$, gives $T_{BKT}=0.709(2)$, and discarding the points for $L<40$ gives $T_{BKT}=0.704(3)$, which agrees very well with previous results~\cite{Evertz,PLA_Costa}, $0.700(5)$ and $0.700(1)$ respectively. More precise results could be obtained by increasing the number of zeros in the map by reducing $\varepsilon$, increasing the precision in the location of the cusp. However, the zeros finder task may become a problem for such a high degree polynomial ($> 3 \times 10^{4}$). In any case, conventional methods to locate the $BKT$ may be used together with this one to get even more precise results for $T_{BKT}$.

\noindent

    From Figs.~\ref{zeros_map_zoom} and~\ref{border_line} one can see that the curves diverge from the positive real axis, in accordance with the expectation that for $T>T_{BKT}$ the imaginary part of the zeros should remain finite. Moreover, since the free energy \emph{has to be} an analytic function at high temperatures, \emph{there can be no real positive zero of the partition function at high temperatures}.
\noindent
\begin{figure}
    \includegraphics[width=0.45\textwidth,keepaspectratio=true,clip=true]{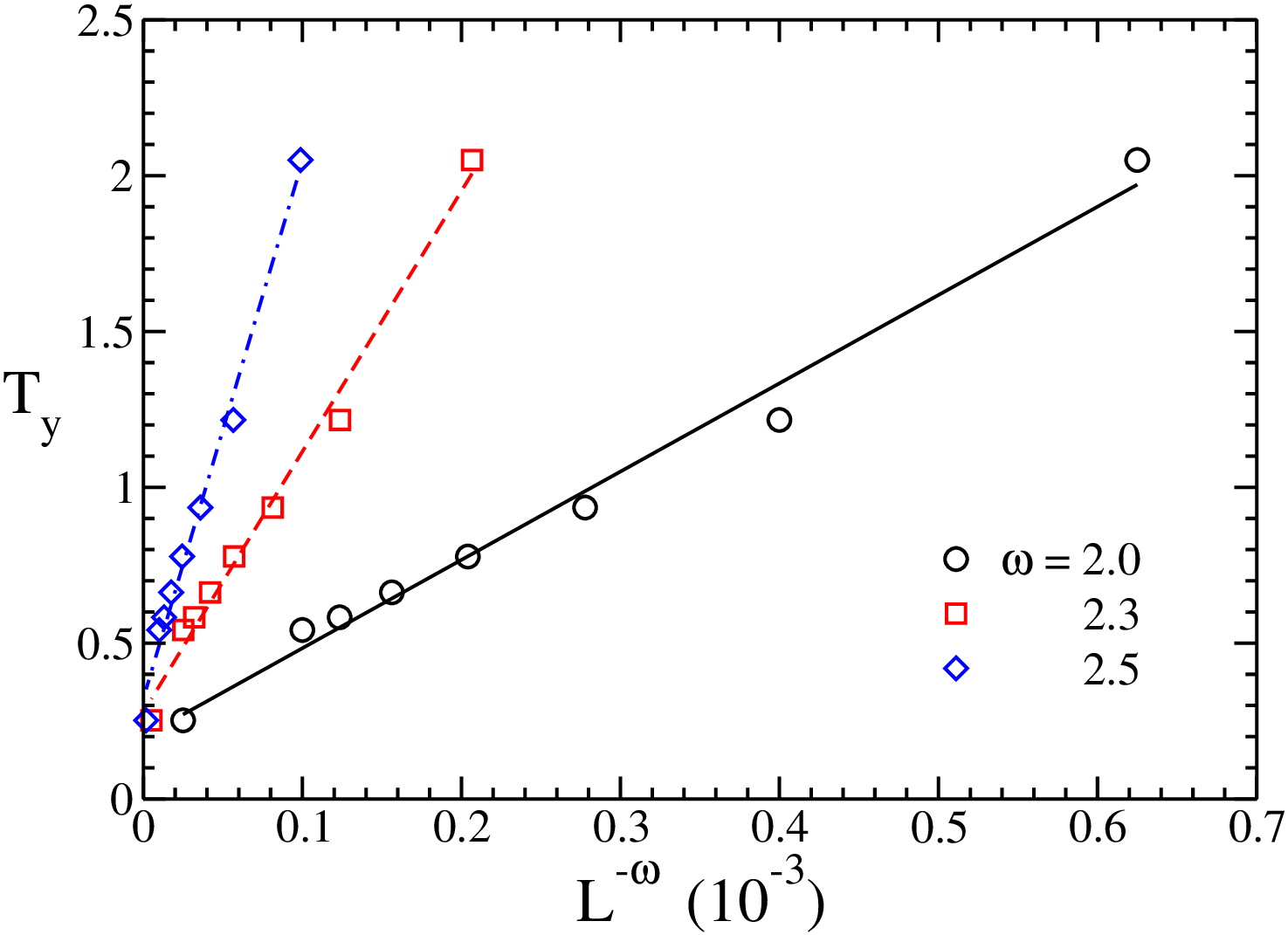}
     \caption{Typical $T_y \times L^{-\omega}$ behavior. Here $T_x =3.0 > T_{BKT}$. Linear fit gives support to a power law behavior with exponent $\omega \approx 2.4$.
    \label{scaling}}
\end{figure}

\noindent

    To analyse the behavior of $T_y$ in the non-critical region we have to be careful. First we have to consider the following. An analysis based in finite size scaling, as we did in the critical region, is meaningless since the basic assumption of any $FSS$ is that the free energy behaves as a homogeneous function. Following Ytzykson we identify three regions. (1) The region far from $T_{BKT}$ where $L >> \xi \approx 1$, where finite size lattice effects are not relevant. Here $\xi$ is the correlation length which for the $XY$ model diverges exponentially when $T_{BKT}$ is approached from above, remaining infinite in the critical region. (2) The region $T \gtrsim T_{BKT}$ where $L >> \xi >> 1$. The finite lattice exhibits the same scaling behavior as the infinite system. And, (3) the region where $L \approx \xi >> 1$ for which are observed severe finite size effects. Regions (2) and (3) are scaling regions.
With that in mind we content ourselves with an analysis in a region far enough from $T_{BKT}$ where we may expect that the border of the zeros map should converge fast to its asymptotic limit.
For $T > T_{BKT}$ the free energy is an analytical function of $T$ so that it can be approximated by a polynomial in $T$ with the coefficients depending on the lattice size in powers of $1/L$. Farthest we are from $T_{BKT}$ faster will decay the higher coefficients of the polynomial,
so that, it is reasonable to suppose that in this region $T_y$ has a power law $T_y(\infty) + AL^{-\omega}$ behavior. Asymptotic convergence is assured by the Brouwer fixed point theorem \cite{Marshall}. Reasoning in this way we can plot $ T_y $ for the largest lattices of our simulations as a function of $L^{-\omega}$. By varying $\omega$ we can find the best value that adjusts a linear function to our data. This procedure is shown in Fig.~\ref{scaling}. It is noteworthy the different ``scaling'' behavior of $\mathfrak{Im}(T)$ in both regions, $T < T_{BKT}$ and $T > T_{BKT}$. As should be expected the exponent $\omega$ is not universal, depending on $T$. 

\section{Conclusion}
    In summary, we show that a method of investigating the Fisher zeros of the partition function can identify whether or not the model exhibits a Berezinskii-Kosterlitz-Thouless (BKT) phase transition. By studying the 2D XY-model we found a qualitative picture that is completely consistent with expectations for the $BKT$ transition, i.e., the zeros map is consistent with the existence of an entire line of zeros in the real positive axis in the thermodynamic limit, with different behaviors for points below and above the transition temperature that could be used to signalize the $BKT$ transition. Moreover, the $BKT$ transition temperature was successfully obtained by considering the location of the cusp that splits regions with different behaviors giving $T_{BKT}=0.704(3)$, in excellent accordance with previous results~\cite{PLA_Costa,Evertz}. Our quantitative analysis shows that for temperatures above $T_{BKT}$ the imaginary part of the zeros converge to finite values. This behavior is in accordance with the fact that the free energy \emph{must be} an analytical function over the real axis at high temperatures, which prevents the existence of any real positive zero at high~$T$.
\section*{Acknowledgments}
    We would like to thank Prof. David P. Landau for  very fruitful discussions and the Center for Simulational Physics at UGA where part of this work was realized. This work was partially supported by CNPq and Fapemig, Brazilian Agencies. JCSR thanks CNPq for the support under grant PDJ No. 150503/2014-8 and Grant CNPq 402091/2012-4.

\section*{References}
\bibliographystyle{elsarticle-num}
\bibliography{Zeros_v5}

\end{document}